# A Blast Wave from the 1843 Eruption of Eta Carinae

Nathan Smith*

*Astronomy Department, University of California, 601 Campbell Hall, Berkeley, CA 94720-3411*

**Very massive stars shed much of their mass in violent precursor eruptions [1] as luminous blue variables (LBVs) [2] before reaching their most likely end as supernovae, but the cause of LBV eruptions is unknown. The 19th century eruption of Eta Carinae, the prototype of these events [3], ejected about 12 solar masses at speeds of 650 km/s, with a kinetic energy of almost $10^{50}$ergs[4]. Some faster material with speeds up to 1000-2000 km/s had previously been reported [5,6,7,8] but its full distribution was unknown. Here I report observations of much faster material with speeds up to 3500-6000 km/s, reaching farther from the star than the fastest material in earlier reports [5]. This fast material roughly doubles the kinetic energy of the 19th century event, and suggests that it released a blast wave now propagating ahead of the massive ejecta. Thus, Eta Carinae's outer shell now mimics a low-energy supernova remnant. The eruption has usually been discussed in terms of an extreme wind driven by the star's luminosity [2,3,9,10], but fast material reported here suggests that it was powered by a deep-seated explosion rivalling a supernova, perhaps triggered by the pulsational pair instability[11]. This may alter interpretations of similar events seen in other galaxies.**

Eta Carinae [3] is the most luminous and the best studied among LBVs [1,2]. It is the most massive and luminous star in our region of the Milky Way, and it provides important constraints on the pre-supernova (SN) evolution of the most massive stars, even if it is a rather extreme example of the instabilities that they encounter. Eta Carinae probably began its life with an initial mass of around 150 $M_{\odot}$, and has a current estimated mass of about 90-100 $M_{\odot}$, with much of the difference lost in sudden giant eruptions in the past few thousand years [1]. It is orbited by a companion star [12], but it is unclear whether the companion played a role in its eruptive instability [4].

It is now well established that the so-called "Homunculus" nebula was ejected during the 1840's near the peak of Eta Carinae's eruption [13]. That event ejected at least 12 $M_{\odot}$ of gas moving at speeds up to 650 km/s [4,14]. This speed roughly matches the present-day polar wind speed [15] and is close to the expected escape velocity from the star's surface. The observed ejecta speeds reported here are much faster than this, and are quite surprising since such high speeds are not expected in a wind from an evolved massive star like Eta Carinae.

The data reported here come in two varieties. First, near-infrared spectra of the He I λ10830 emission line obtained with the GNIRS instrument on the Gemini South telescope (Fig. 1) reveal emission from what appears to be a fast out-flowing disk or material near the waist of bipolar lobes outside the Homunculus, with a spatial scale and expansion speeds roughly twice those of the well-known equatorial skirt of the Homunculus. Although extended He I λ10830 emission had been discovered at one position outside the Homunculus in a previous study [6], the larger structure and faster velocities reported here have not been seen before in any data. This material is flowing away from the star with radial speeds of roughly 1000-2000 km/s, and appears to be present all the way around the Homunculus not far from the equatorial plane.

Second, visual-wavelength spectra of the region near Hα show [N II] λ6548 and λ6583 emission from extremely fast material with Doppler shifts of roughly -3000 to +2500 km/s (see Fig. 2 and Fig. S1 in the Supplementary Information). Emission from this fast material was suspected to exist in previous data of lower quality obtained at one position around Eta Carinae [5], and provided the main motivation to obtain the high-quality data seen here. The fast N-rich gas reported here is spatially coincident with or inside the soft X-ray shell around Eta Carinae (Fig. S1). It is apparently running into the many N-rich outer knots seen in visual-wavelength images [5,16,17,18], because it is near or interior to these knots, but is expanding outward at a much faster pace. The fastest material has not been seen in most [N II] images of Eta Carinae, like those obtained with the *Hubble Space Telescope* (Fig. 1), because it is Doppler shifted far out of the narrow filter bandpass. It follows the same bipolar expansion pattern seen for the Homunculus [14], as well as that of the slower [N II]-bright knots in the outer ejecta [19] (blueshifted to the south-east, redshifted to the northwest). The observed Doppler shifts of up to 3000 km/s are a lower limit to the deprojected maximum velocity of this material. Since this material is seen far off to the side of Eta Carinae, its trajectory must be inclined away from our line of sight, with plausible values for the true space velocity ranging from 3500—6000 km/s (Fig. S2 in the Supplementary Information). The higher speeds would require that this material has been accelerated by the pressure behind the blast wave since its ejection.

The fastest material has [N II]/Hα ratios similar to the slower N-rich knots [18], indicating that it was launched at these speeds by the same evolved primary star that ejected the N-rich Homunculus [21] and all its other N-rich material. While it is intriguing that speeds around 3,000 km/s are similar to the wind speeds of the secondary star in models for the X-ray emission from Eta Carinae [22], the notion that perhaps the secondary star was responsible for ejecting this fast N-rich material has little merit upon closer examination. If one ascribes the 1843 outburst to the as yet unseen secondary star in the Eta Carinae binary system, several inconsistencies arise: First, the expansion speed, bipolar shape, and polar axis orientation of the Homunculus [14] (known to be ejected in the 1843 event [13]) match the present-day properties of the primary star's wind [15]. Second, the star we observe now as the primary is wildly variable, has an enormous mass-loss rate, and shows visible signs of recovery from the 1843 eruption, while the putative secondary star has not been detected. Third, for an evolved massive star to produce the wind speeds required of the secondary star, it must be a compact H-poor Wolf-Rayet star, but this is incompatible with Eta Car's multiple ejections of massive H-rich nebulosity. In short, passing the burden of mass ejection to Eta Car's companion star is not viable.

In any case, speeds of this order evoke properties more like SN explosions than super-Eddington winds [9], and they provide a fundamental new clue to the nature of the trigger behind Eta Car's eruption. Combining the fast near-equatorial material traced by He I λ10830 with fast polar ejecta traced by [N II] implies a bipolar forward shock geometry, perhaps like that depicted in Figure 3. The geometry is similar to that of the Homunculus itself [7,14], but 3-4 times its size and speed. This fast blast wave from the 19th century eruption of Eta Carinae is powered by the very fast and low density N-rich material that is coincident with or interior to the X-ray shell [23]. Seward et al. [23] found that the properties of the soft X-ray shell were consistent with a blast-wave interpretation. It is the interaction between the blast wave from the 1843 event overtaking the 500-1000 yr old [17] clumpy N-rich ejecta shell – and *not* the older condensations running into ambient material – which apparently gives rise to the X-ray emission and shock ionization of the outer ejecta. This fast blast wave is likely to be important for the total energy budget of the 1843 eruption. If the high speed material contains only 5% of the mass of the Homunculus, for example (corresponding to a likely average density of at least 500 cm$^{-3}$ filling the volume of a sphere over the range of radii from the star where the fast material is seen with a ~0.5 filling factor), but is moving at speeds 5 times faster, it may contain $7\times10^{49}$ ergs. This amount of kinetic energy would be comparable to or could even exceed the kinetic energy of the main Homunculus nebula [4].

What could be the origin of this extremely fast and energetic forward shock? Currently favoured models for the mass loss during a giant LBV eruption involve a radiation-driven wind as the star exceeds the Eddington luminosity limit [2,3,9,10], but in that case one expects a wind to be quite slow [9]. Fast material may escape through regions between dense clumps if the wind becomes highly porous [9], but this still doesn't explain how material is accelerated to speeds far exceeding the star's escape speed. Instead, an event of this sort, which also launched the 12 $M_\odot$ of matter in the Homunculus [4], may have been a deep-seated explosion. Its constraints are reminiscent of the pulsational pair instability [11] or some other instability associated with explosive nuclear burning in the latest stages of evolution. However, those burning events are expected to occur shortly (~10 yr) before the final core collapse SN explosion, as hypothesized for the progenitor of the extremely luminous SN 2006gy [24,25], or in some cases as much as 1,000 yr before core collapse. In the unique case of SN 2006jc, a precursor LBV-like outburst was actually observed just 2 yr prior to the final supernova [26,27]. Eta Car has not yet reached core collapse even though the eruption happened more than 160 yr ago, and it is thought to have undergone similar eruptive events ~1000 yr ago [17].

Altogether, the presence of a fast blast wave has critical implications for understanding the nature of Eta Carinae's 1843 eruption, and the LBV instability in general. In effect, it requires that these outbursts can have an explosive aspect that has not previously been appreciated, blurring the observational and phenomenological distinctions between giant LBV eruptions and SN explosions in the most massive stars. A corollary is that the extended nebula now seen around Eta Carinae is analogous to a low-energy supernova remnant, and the X-ray emission arising from the interaction between the blast wave and surrounding material may have been stronger in the past. Eta Carinae is also the prototype for a class of stellar eruptions seen in other galaxies, sometimes called "SN impostors", Type V supernovae, or faint Type IIn supernovae [28]. Indeed, there may even be an ill-defined continuum in energy between eruptions like Eta Carinae, super-outbursts of LBVs like SN 1961V [29], and faint core-collapse SNe. Since these LBV outbursts can evidently have powerful blast waves of a few thousand km/s, the observations reported here indicate that detections of radio synchrotron or X-ray emission are *not* reliable ways to distinguish between LBV eruptions and genuine core-collapse SN explosions.

**Supplementary Information** accompanies the paper on **www.nature.com/nature**.

**Acknowledgements** The author acknowledges ongoing collaboration and relevant discussions with the supernova group at UC Berkeley.

**Author Information** Correspondence and requests for materials should be addressed to N.S. (e-mail: nathans@astro.berkeley.edu).

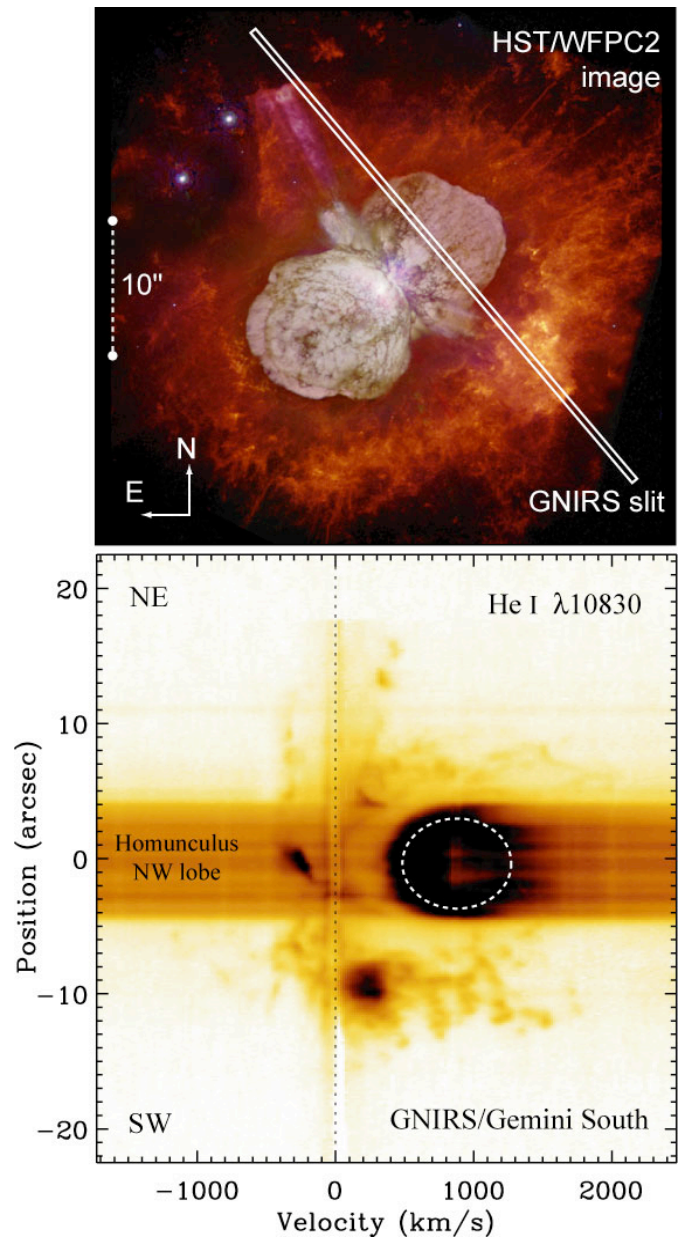

**Figure 1. Example of the velocity structures seen in the He I λ10830 line.** The 25x0.3 arcsec spectroscopic slit aperture of the GNIRS spectrograph was oriented at P.A.=+40 deg (SW to NE) crossing the NW polar lobe of the Homunculus nebula (see image; top). In the resulting position-velocity plot (bottom), the broad horizontal strip is from reflected continuum light scattered by dust in the Homunculus NW polar lobe, while He I λ10830 emission from fast material is seen outside of that structure over a large range of velocities. The top image is a composite-colour visual wavelength image made from several different exposures obtained using the WFPC2 camera on the *Hubble Space Telescope*, through the F336W (blue), F631N (green), and F658N (red) filters, and is shown here only for a comparison of the ejecta morphology with the slit placement. These data were reduced in a manner consistent with previous similar data [6,14]. The outer gas that emits He I in the bottom panel spectrum appears

red/orange in this image because of strong [N II] emission. The position-velocity plot (lower panel) is a sub-section of a 1.0-1.15 μm spectrum centred on the bright He I λ10830 emission line, obtained on 2005 March 20 using the GNIRS spectrograph mounted on the Gemini South telescope. It is a combination of four individual exposures of 90 s each. The image (top) and long-slit spectrum (bottom) have the same spatial scale. The dashed white ellipse in the bottom panel marks a He I feature that arises from emission formed in the central star's wind, but is scattered and red-shifted by dust grains; it is unrelated to the current topic.

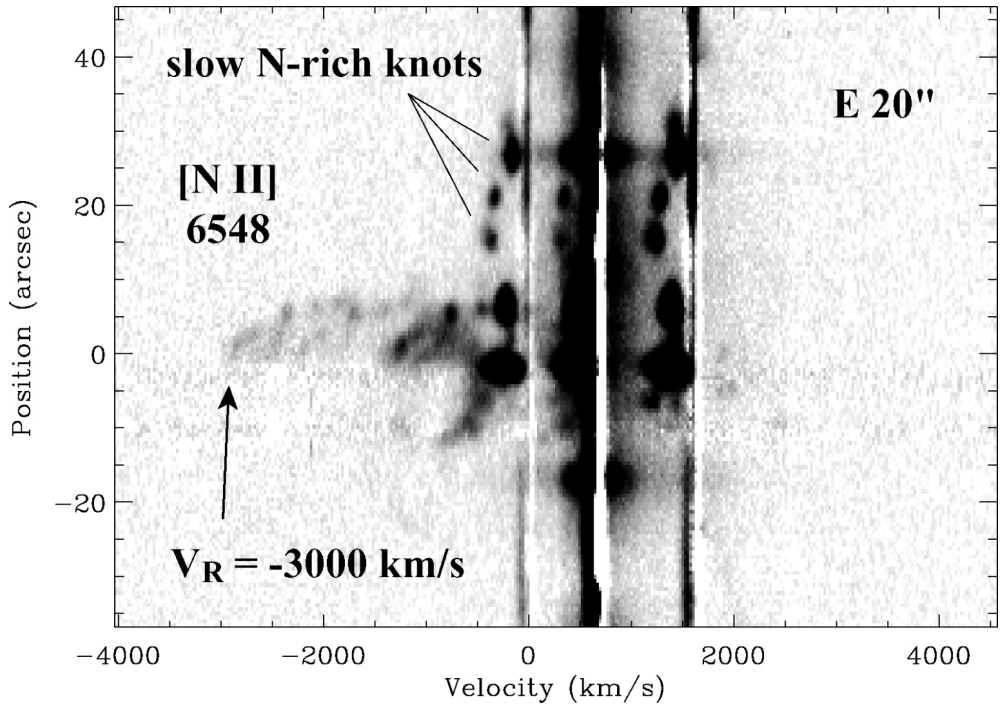

**Figure 2. Extremely fast nitrogen-rich material surrounding Eta Carinae.** This panel shows a 2-D position/velocity spectrum at an example offset position 20-arcsec east of the star, resulting from a 15-min exposure obtained on 2006 March 15 using the RC- Spec spectrograph mounted on the CTIO 4-m telescope. Similar visual-wavelength spectra at several additional offset positions are included in the Supplementary Information (Fig. S1). The velocity scale is set for the [N II] λ6548 line to aid the interpretation, because blue-shifted emission dominates at this position. [N II] λ6548 emission reveals material travelling at blue-shifted Doppler speeds up to 3000 km/s, coincident with the outer shell seen in X-rays, while the dense N-rich knots seen in *HST* images have much slower speeds of 100-300 km/s. The slow-moving knots have been studied in detail [5,16,17,18,19], but the faster material at -2000 to -3000 km/s had only been suspected at one location in a single exposure in previous data of lower quality [5]. These earlier observations provided the motivation for the higher-quality spectra presented here.

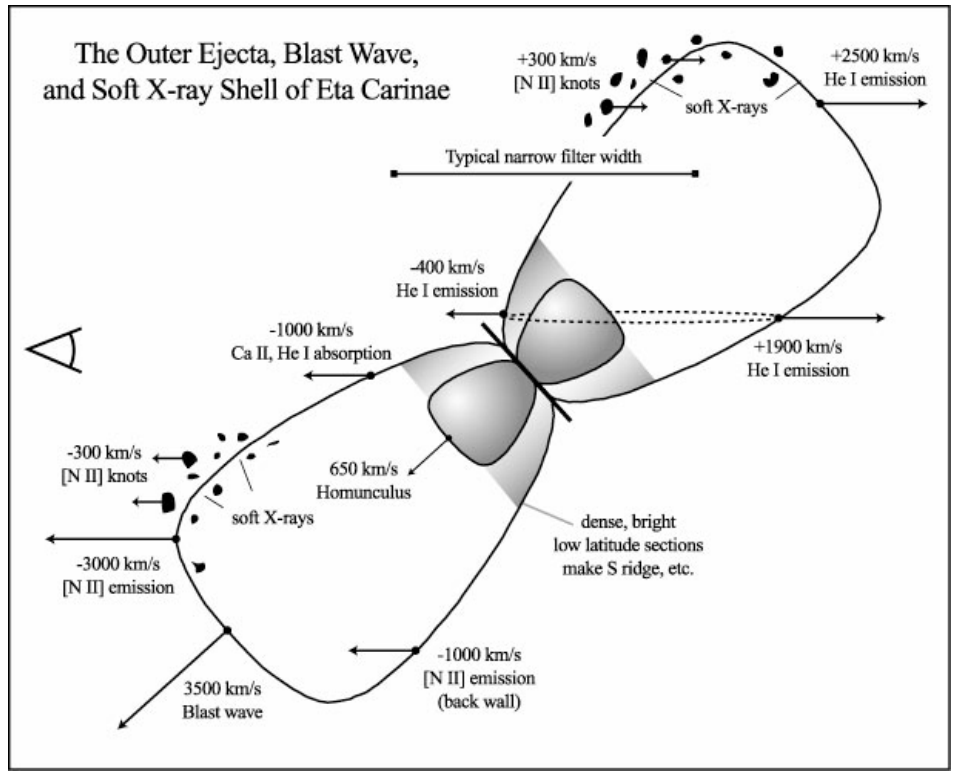

**Figure 3. Illustration of the possible geometry of Eta Carinae's blast wave.** This simplified diagram depicts the basic geometry of Eta Carinae's outer ejecta, with an Earth-based observer located to the left. The inner shaded structure shows the shape and orientation of the Homunculus nebula [7,14]. Some of the relatively slow, dense N-rich knots from a previous eruptive event ~1000 yr ago [17] are now being overrun by the fast forward shock from the 1843 eruption of Eta Car. This collision gives rise to the soft X-ray shell [23], and UV radiation from shocked gas ionizes pre-shock material in the slower N-rich knots seen in images. For background information on previous studies of these slower outer debris, see refs [5,16,17,18,19,20]. Part of the thin side wall of the approaching side of this bipolar shock structure crosses in front of our line- of-sight to the Homunculus nebula, which is a reflection nebula. This part of the structure, blue-shifted at roughly -1000 km/s, may have been seen previously in absorption in the Ca II HK [7] and He I λ10830 [6] lines, and in emission in Hα and [N II] lines [7,8], but the connection between these velocity features and the polar blast wave shown here is not entirely clear with available data. A typical narrow-band imaging filter, like the F658N filter of the WFPC2 camera on *HST*, will exclude the fastest Doppler-shifted features.

## Supplementary Information

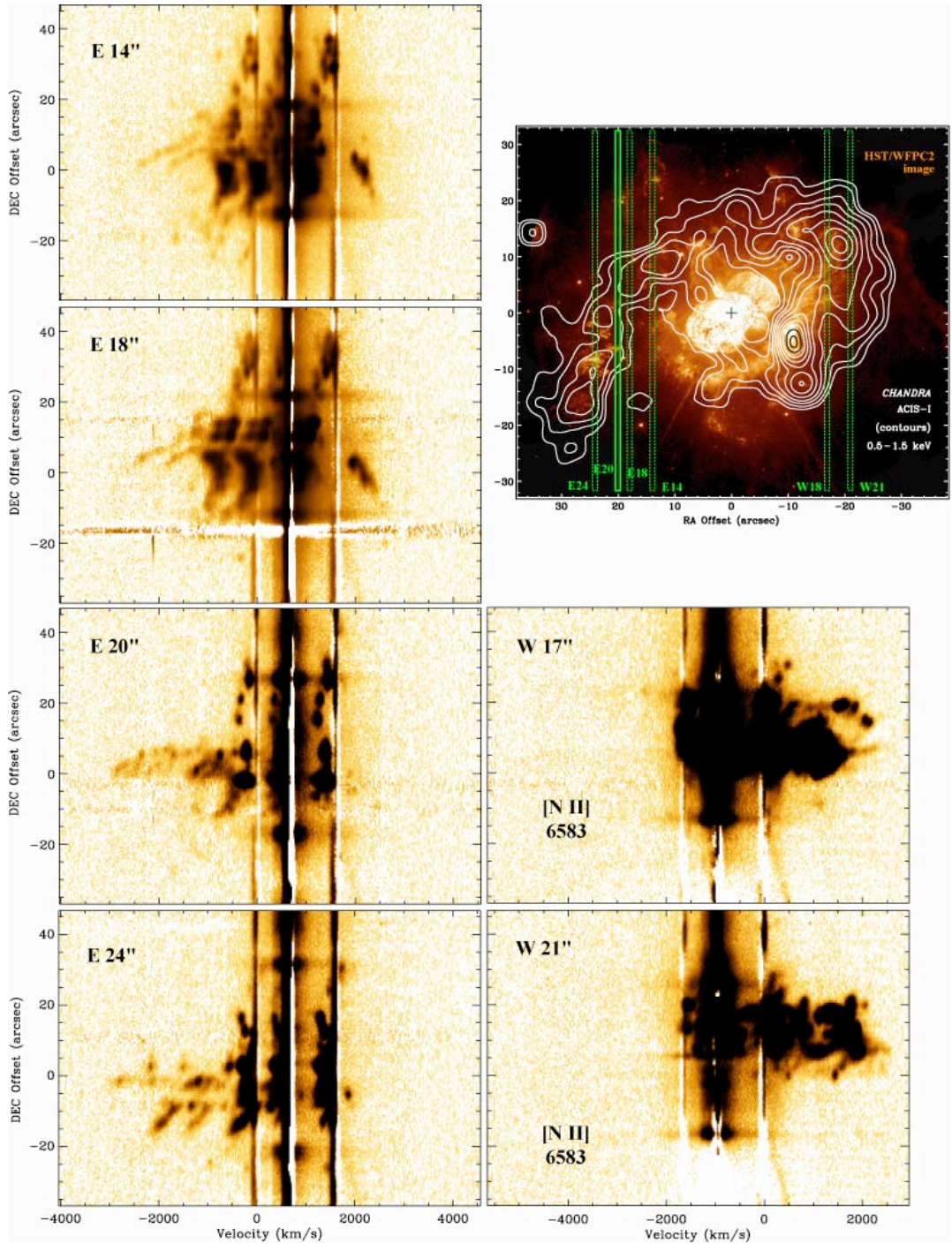

**Supplementary Figure S1.** The top right panel shows a *Hubble Space Telescope* (*HST*) image obtained with the WFPC2 camera's WF3 chip using the F658N filter, tracing [N II] λ6583 emission [see ref. 13]. Superposed are contours of soft X-ray emission obtained with the Chandra X-ray Observatory [23], as well as slit positions of visual-wavelength spectra at various offsets. The image and contours are taken from [5], and are reproduced here for comparison with the spectral features. The remaining panels are the same as in Figure 2, but for all 6 offset slit aperture positions to the east (left) and west (right) of the star noted in the image (these data were obtained on the same night and with the same instrument and telescope setup as the spectra in Fig. 2). Position-velocity plots on the left column are made to show the Doppler shift relative to the shorter-wavelength [N II] λ6548 line, to emphasize the fastest blue-shifted material. Likewise, the two panels in the right column are plotted for the longer-wavelength emission line in the doublet, [N II] λ6583, in order to emphasize the fastest red-shifted material. The fastest blue-shifted features appear to outline the thin outer part of a polar cavity, perhaps the approaching side of a bipolar blast wave as depicted in Figure 3, especially in the offset positions 14" and 18" east, which slice through the interior parts of the cavity.

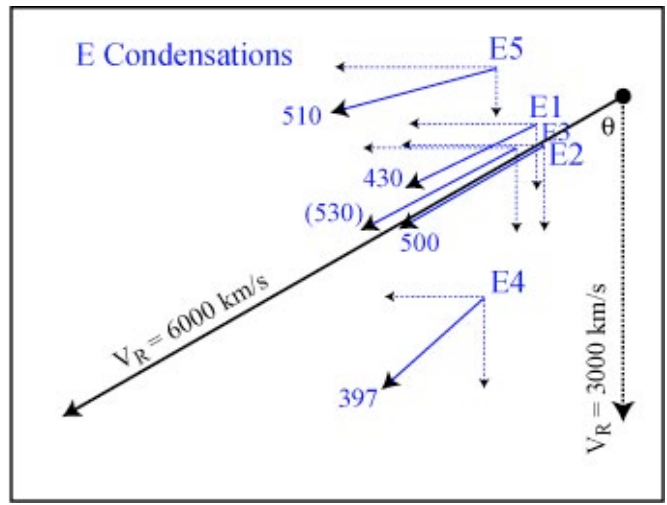

**Supplementary Figure S2.** Trajectories (blue) of the dense and relatively slow "E Condensations" [16], the group of dense knots located 20-24 arcsec east of the star in the image in Figure S1. Line-of-sight radial velocities measured in our spectra are combined with proper motion measurements [17] to derive true space velocities (values noted in blue in km/s) and trajectory angles from the plane of the sky ranging from 15 deg (E5) to 42 deg (E4). The plane of the sky is horizontal in this figure, and the observer is at the bottom. If the fast material with Doppler shifts of -3000 km/s is directed at an angle of 30 deg from the plane of the sky, shown here, similar to E1, 2, and 3, then its true space velocity is roughly 6,000 km/s. If on the other hand it is directed closer to our line of sight and tilted about 60 deg from the plane of the sky, appropriate for linear motion and an age of 160 yr (placing it near the polar axis of the Homunculus), then its true space velocity is about 3500 km/s.